\documentclass[aps,prl,twocolumn,superscriptaddress]{revtex4}
\usepackage{graphicx}
\usepackage{color}

\begin{document}
\title{Electronic Structure of Epitaxial Single-Layer MoS$_2$}
\author{Jill A. Miwa}
\author{S\o ren Ulstrup}
\author{Signe G. S\o rensen}
\author{Maciej Dendzik}
\author{Antonija Grubi\v{s}i\'c \v{C}abo}
\author{Marco Bianchi}
\author{Jeppe Vang Lauritsen}
\author{Philip Hofmann}
\affiliation{Department of Physics and Astronomy, Interdisciplinary Nanoscience Center (iNANO), Aarhus University, 8000 Aarhus C, Denmark}
\email[]{philip@phys.au.dk}

\date{\today}
\begin{abstract}
The electronic structure of epitaxial single-layer MoS$_2$ on Au(111) is investigated by angle-resolved photoemission spectroscopy. Pristine and potassium-doped layers are studied in order to gain access to the conduction band. The potassium-doped layer is found to have a (1.39$\pm$0.05)~eV direct band gap at $\bar{K}$ with the valence band top at $\bar{\Gamma}$ having a significantly higher binding energy than at $\bar{K}$.  The moir\'e superstructure of the epitaxial system does not lead to the presence of observable replica bands or minigaps. The degeneracy of the upper valence band at $\bar{K}$ is found to be lifted by the spin-orbit interaction, leading to a splitting of (145$\pm$4)~meV. This splitting is anisotropic and in excellent agreement with recent calculations. Finally, it is shown that the strength of the potassium doping is $k$-dependent, leading to the possibility of band structure engineering in single-layers of transition metal dichalcogenides. 
\end{abstract}
\pacs{73.22.-f,73.20.At,79.60.-i}

\maketitle

Soon after the first isolation of graphene \cite{Novoselov:2004,Novoselov:2005,Zhang:2005}, it became clear that other layered materials could also be thinned down to a single layer using the same methods, and that such layers may have interesting properties \cite{Novoselov:2005b}. Particular focus has been on MoS$_2$ \cite{Mak:2010,Splendiani:2010,Cao:2012b,Radisavljevic:2011,Zeng:2012,Xiao:2012b}, a material that had been grown in single layers and used in catalysis even before the advent of graphene \cite{Topsoe:1996,Helveg:2000,Lauritsen:2003,Jaramillo:2007}. Single-layer (SL) MoS$_2$ has indeed  been shown to have a number of intriguing properties. To name but a few, SL MoS$_2$ has a direct band gap  in contrast to the bulk \cite{Bollinger:2001} and correspondingly different optical properties \cite{Mak:2010,Splendiani:2010}. Having the conduction band minimum at the $\bar{K}$ point of the Brillouin zone (BZ) opens interesting possibilities for new valley and spin-valley physics \cite{Cao:2012b,Zeng:2012,Xiao:2012b}. It is also possible to construct transistors based on SL MoS$_2$ with the advantage of a high on/off ratio compared to (bilayer) graphene-based devices \cite{Radisavljevic:2011}. 

Instead of obtaining SL MoS$_2$ by micro mechanical exfoliation, high-quality layers can be grown on different substrates, enabling a new avenue for fundamental investigations of this material. While the growth of nano scale SL MoS$_2$ clusters is particularly well established \cite{Helveg:2000,Lauritsen:2007}, it has recently become possible to grow large area epitaxial SL MoS$_2$ \cite{Sorensen:2014}. Similar to many epitaxial graphene systems \cite{Berger:2004,Diaye:2008}, the structure of SL MoS$_2$ shows a strong moir\'e superlattice due to the lattice mismatch with the underlying Au(111) \cite{Sorensen:2014}. In this work, we exploit the very high quality and large areas obtainable for epitaxial SL MoS$_2$ to study its electronic structure by angle-resolved photoemission spectroscopy (ARPES). This provides a detailed picture of the new effects arising from quantum confinement, breaking of the bulk inversion symmetry, the role of  spin-orbit coupling, as well as the effect of the underlying Au(111) and the moir\'e structure.

Epitaxial SL MoS$_2$ has been grown on Au(111) by methods described elsewhere \cite{Sorensen:2014}. Actually, the total MoS$_2$ coverage used here was kept somewhat below one monolayer at $\approx 0.65$~ML in order to avoid the growth of 2~ML islands that were found to be detectable in ARPES. The epitaxial SL MoS$_2$ samples are stable in air and could thus be removed from the dedicated growth chamber, transported to the SGM-3 end station on the synchrotron radiation source ASTRID2 \cite{Hoffmann:2004} and cleaned via mild annealing to 500~K, a procedure that has been verified to yield atomically clean surfaces by scanning tunnelling microscopy (STM). ARPES data were collected at 80~K with an energy and angular resolution better than 20~meV and 0.2$^{\circ}$, respectively. All measurements presented here were performed with a photon energy of 49~eV. Even though the band structure of SL MoS$_2$ is easily distinguished from the photoemission features from the underlying Au(111), photon energy scans were performed to confirm the lack of $k_z$ dispersion of the SL MoS$_2$ bands.

\begin{figure*}
\includegraphics[width=0.8\textwidth]{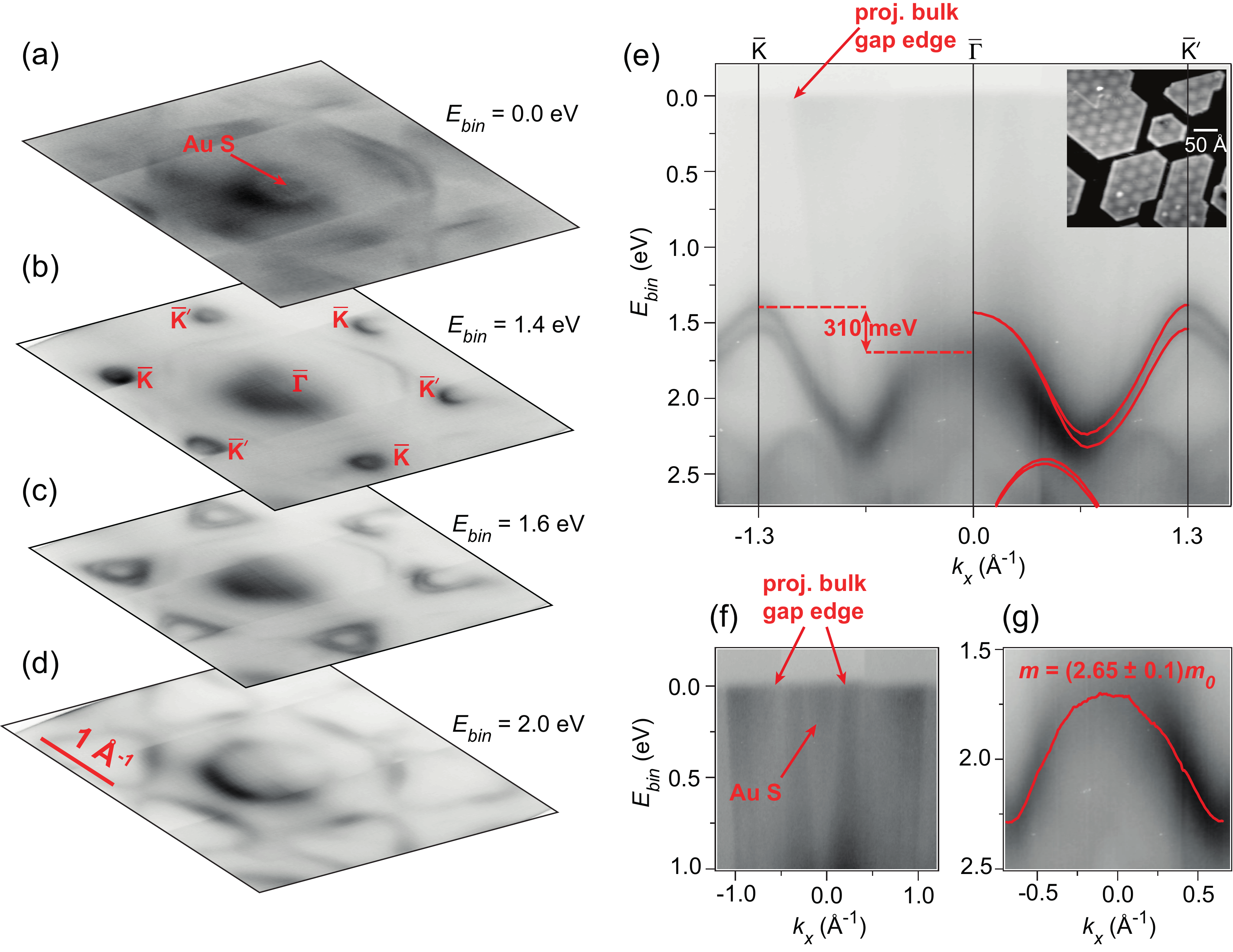}\\
\caption{(Color online) Electronic structure of epitaxial single-layer MoS$_2$: (a)-(d) Constant energy slices through the first Brillouin zone, showing (a) the Fermi contour dominated by Au bulk states and the surface state (Au S) and (b)-(d) evolution of MoS$_2$ valence band features around $\bar{K}$, $\bar{K}^{\prime}$ and $\bar{\Gamma}$ points. (e) Valence band dispersion in the $\bar{K}-\bar{\Gamma}-\bar{K}^{\prime}$ direction. The band structure for free-standing single-layer MoS$_2$ from Ref. \onlinecite{Zhu:2011d} has been superimposed. The inset shows an STM image of the MoS$_2$ islands and the moir\'e ($I_t$=0.4~nA, $V_t$=-1.2~V).  (f) Close-up of dispersion close to the Fermi level, dominated by Au features including a broad Au surface state around $\bar{\Gamma}$. (g) Close-up of MoS$_2$ upper valence band dispersion around $\bar{\Gamma}$. The red line is the peak position extracted  from energy distribution curves. The curvature of the parabolic band provides the stated effective mass $m$ in units of the free electron mass $m_0$. The two-dimensional high symmetry points in this figure refer to the SL MoS$_2$ structure.}
  \label{fig:1}
\end{figure*}

Figure \ref{fig:1} gives an overview of the epitaxial SL MoS$_2$ band structure.  The constant binding energy cuts in Fig. \ref{fig:1}(a-d) reveal both Au(111) and SL MoS$_2$ features. The Au(111)-related states are best identified near the Fermi energy due to the lack of SL MoS$_2$ states there. 
At higher binding energies, maxima in the upper valence band (VB) states of SL MoS$_2$ are observed, both at the $\bar{\Gamma}$ as well as at the $\bar{K}$ points. These features are also seen in the measured dispersion shown in Fig. \ref{fig:1}(e). While the states near $\bar{K}$ are very distinct and sharp, those near $\bar{\Gamma}$ are rather broad. This is ascribed to the different interaction with the substrate and the orbital character of the states. The upper VB near $\bar{K}$ falls into a projected band gap of the Au(111) electronic structure \cite{Takeuchi:1991} and can therefore not interact with the bulk states. Indeed, the presence of this gap is even visible in the data of Fig. \ref{fig:1}(e) as a reduction of background intensity between the Fermi energy and $\approx$~2.2~eV binding energy around $\bar{K}$.  Moreover, the upper VB near $\bar{K}$ is derived from in-plane d- and p-orbitals orbitals \cite{Zhu:2011d,Cappelluti:2013} and thus a weak adsorbate-substrate interaction is expected. The upper VB states near $\bar{\Gamma}$, on the other hand, fall within the continuum of projected bulk states \cite{Takeuchi:1991} and are  derived from out-of-plane orbitals \cite{Zhu:2011d,Cappelluti:2013}. For these states a stronger adsorbate-substrate interaction can be expected and this can explain the broadening of the band.

Such an increased interaction is also supported by a comparison of the measured dispersion and the density functional theory band structure for free-standing SL MoS$_2$ by Zhu \emph{et al.} \cite{Zhu:2011d}. In Fig. \ref{fig:1}(e) this calculation is superimposed on the data and aligned at the valence band maximum (VBM) at $\bar{K}$. In the calculation, the upper VB maxima at $\bar{K}$ and $\bar{\Gamma}$ are found at nearly the same binding energy. This is also the case for calculations that include many-body effects \cite{Cheiwchanchamnangij:2012} and in ARPES results from exfoliated SL MoS$_2$ \cite{Jin:2013}. Our data, in contrast, show a distortion of the upper VB with the measured maximum at $\bar{\Gamma}$ being 0.31~eV  lower than at $\bar{K}$.  The distortion has only a small effect on the effective mass near $\bar{\Gamma}$. A fit to a hole-like parabola (see Fig. \ref{fig:1}(g)) gives an effective hole mass of  (2.7$\pm$0.1) times the free electron mass $m_0$, in agreement with the calculation for SL MoS$_2$ (2.8~$m_0$) \cite{Peelaers:2012} and the result for exfoliated SL MoS$_2$ ((2.4$\pm$0.3)~$m_0$) \cite{Jin:2013}, but much higher than the calculated bulk value (0.62~$m_0$) \cite{Peelaers:2012}. 

Figure \ref{fig:1}(f) shows a magnification of the dispersion around $\bar{\Gamma}$ near the Fermi energy. Here the diffuse background intensity is higher in the projected bulk state continuum of Au(111) than in the projected band gaps. The bulk band structure gap opening around the bulk L point leads to a small projected band gap around $\bar{\Gamma}$ \cite{Takeuchi:1991} that is also observed here. Within this gap, we even find a diffuse intensity that is assigned to the well-known surface state on Au(111)  \cite{Reinert:2001}. The corresponding disc of intensity is also visible at the Fermi energy cut in Fig. \ref{fig:1}(a). The surface state's presence under SL MoS$_2$ and also under MoS$_2$ films completely covering the surface (not shown) suggest that the adsorbate-substrate interaction is predominantly of van der Waals character, as in the case of graphene on Ir(111) where a similar phenomenon is observed \cite{Varykhalov:2012b}.   

Another expected consequence of the SL MoS$_2$-substrate interaction would be a manifestation of the pronounced moir\'e in the electronic structure. A  scanning tunnelling microscopy image of the moir\'e is shown in the inset of Fig. \ref{fig:1}(e). A similar moir\'e has pronounced consequences for the electronic structure of  epitaxial graphene, leading to the presence of replica bands and mini-gaps in the Dirac cone \cite{Bostwick:2007,Pletikosic:2009}. Here, such replicas would be expected at a distance of $\approx 0.17$~\AA$^{-1}$~from the main features. We should easily be able to resolve such features, especially for the sharp bands near $\bar{K}$, but we find them absent from the data, suggesting that the electronic structure of SL MoS$_2$ is hardly affected by the moir\'e. We ascribe the difference to graphene to the different character of the states near $\bar{K}$: In graphene, the buckling of the layer directly affects the local interaction of the out-of-plane $\pi$ orbitals with the substrate. In MoS$_2$, on the other hand, the bands have a mix of Mo $d_{x^2-y^2}$, $d_{xy}$ and S $p_x$,$p_y$ character and are thus totally in-plane. Their local interaction with the substrate can be expected to be less affected by the buckling.

\begin{figure}
\includegraphics[width=.49\textwidth]{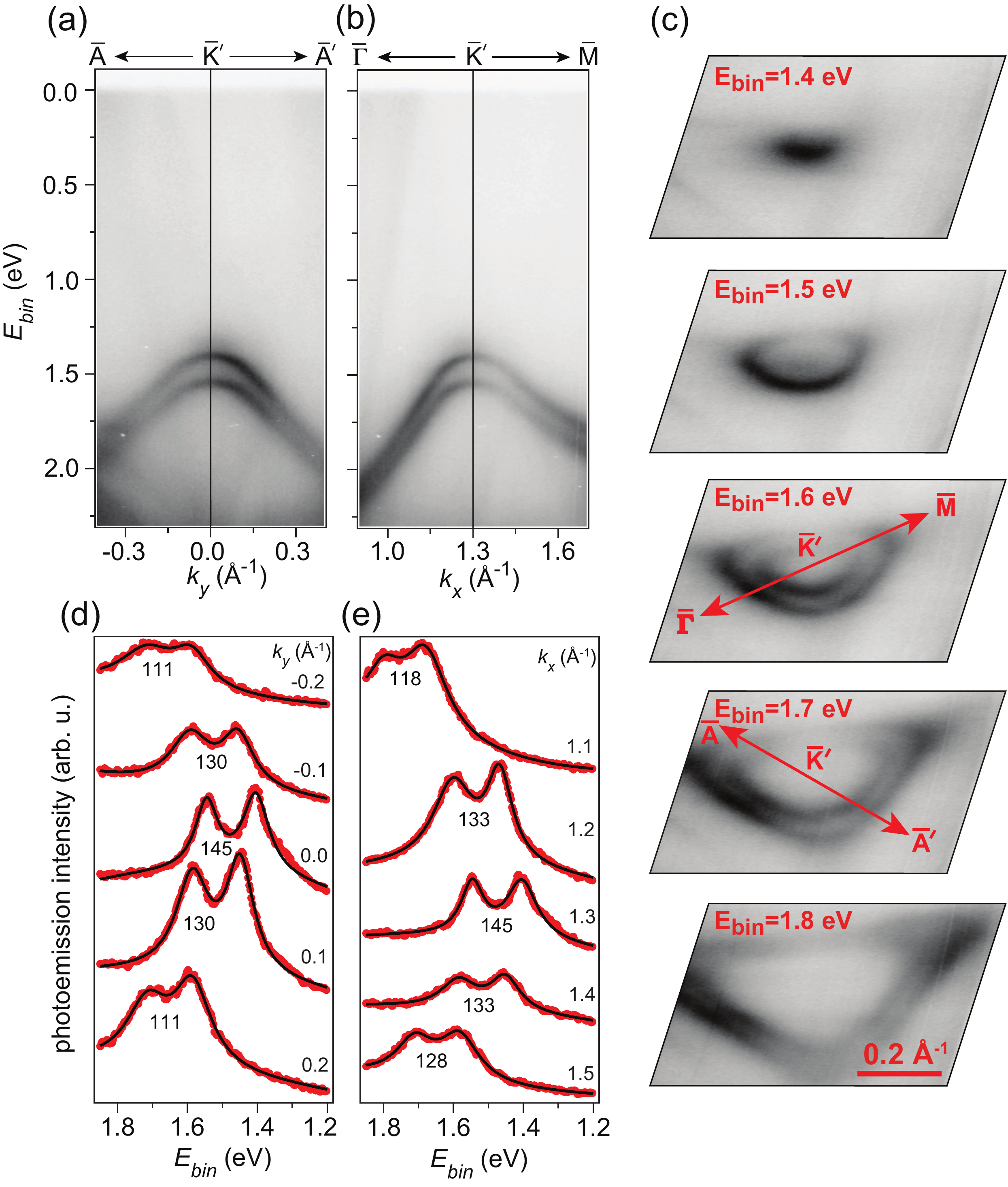}\\
\caption{(Color online) Detailed dispersion around valence band maximum at $\bar{K}^{\prime}$ ($\bar{K}$) and analysis of spin-orbit interaction: (a)-(b) Cuts along directions defined in the constant energy contours in (c). The points $\bar{A}$ and $\bar{A}'$ are along the line perpendicular to the $\bar{\Gamma}-\bar{K}$ direction passing through $\bar{K}$  (d)-(e) EDCs at the given momentum values. The peaks are fitted by a double Lorentzian function, and the difference between peak positions is taken as the  splitting of the bands, which is stated in meV below the curves.}
  \label{fig:2}
\end{figure}

A remarkable effect is the strong spin-orbit splitting of the upper VB near $\bar{K}$, shown in greater detail in Fig. \ref{fig:2}. Note that the splitting in SL MoS$_2$ is a genuine lifting of the spin-degeneracy and different from the splitting in the inversion-symmetric bulk material, where it is a combination of layer interaction and spin-orbit coupling and does not remove the spin degeneracy \cite{Cheiwchanchamnangij:2012}. An equivalent splitting has been observed in ARPES from ML MoSe$_2$ grown on epitaxial graphene \cite{Zhang:2013f} but it has so far remained unresolved for exfoliated SL MoS$_2$ \cite{Jin:2013}. The size of the splitting can be determined from a fit of the energy distribution curves (EDCs) obtained from the data in Fig. \ref{fig:2}(a,b) and shown in Fig. \ref{fig:2}(d,e). The strongest splitting at $\bar{K}$ is found to be (145$\pm$4)~meV. This is somewhat bigger than the value of $\approx 100$~meV obtained by triply resonant Raman scattering \cite{Sun:2013} and, as expected, smaller than the ARPES result for ML MoSe$_2$ of 180~meV. It is in excellent agreement with the theoretical prediction of 148~meV from density functional theory \cite{Zhu:2011d} and 146~meV from GW calculations \cite{Cheiwchanchamnangij:2012}. The anisotropy of the splitting away from $\bar{K}$ that gives rise to a triangular warping of the constant energy contours in Fig. \ref{fig:2}(c) also agrees with the theoretical prediction \cite{Zhu:2011d}.

\begin{figure}
\includegraphics[width=.49\textwidth]{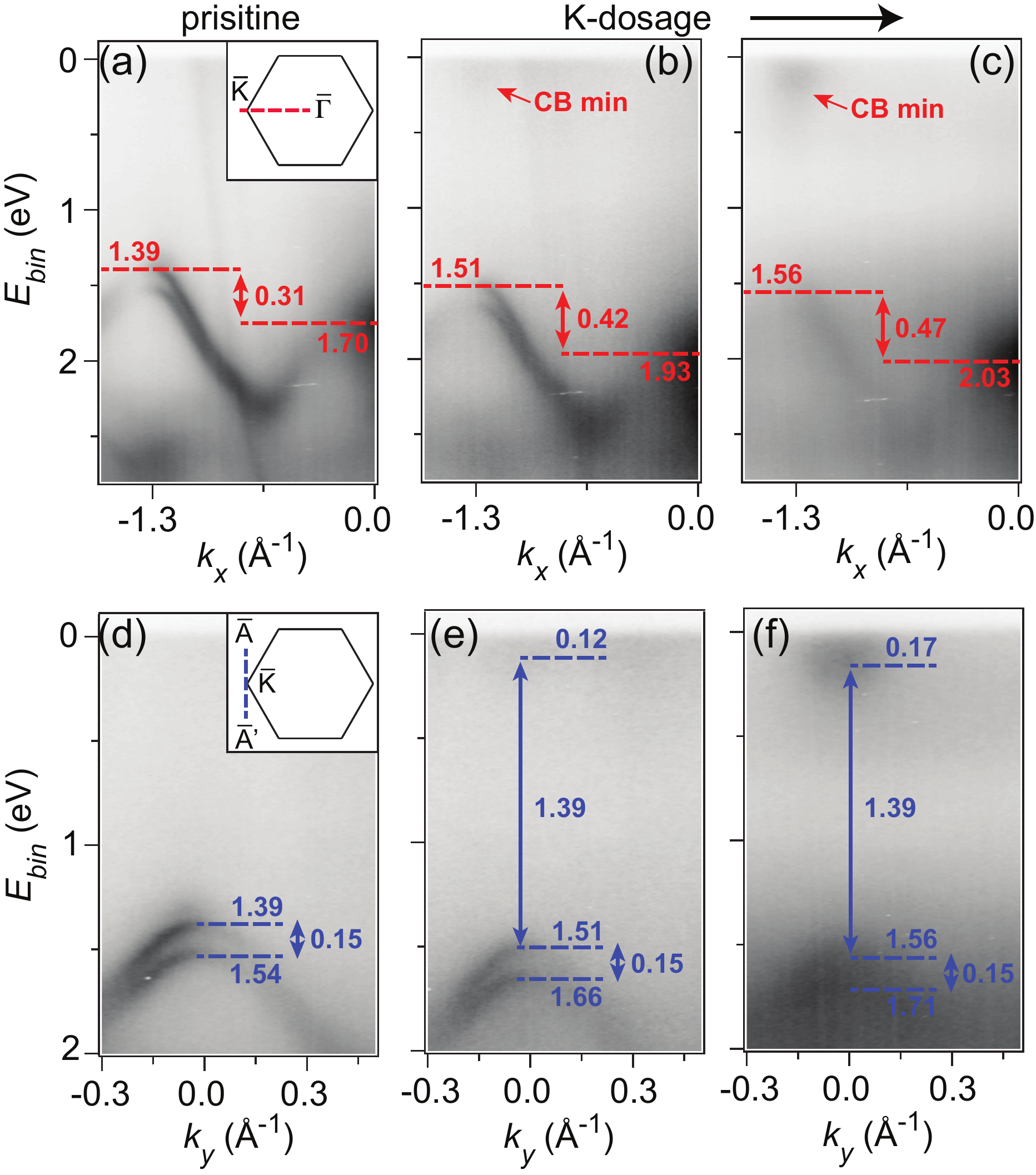}\\
\caption{(Color online) Tuning of the band structure by potassium adsorption: (a), (d) Clean sample, (b),(e) first dose and (c),(f) second dose. The scan directions for the cuts in (a-c) and (d-f) are given in the insets of (a) and (d), respectively. The points $\bar{A}$ and $\bar{A}'$ are defined in the caption of Fig. \ref{fig:2}. The energies are given in eV.}
  \label{fig:3}
\end{figure}

SL MoS$_2$ is expected to be a semiconductor with a direct band gap at $\bar{K}$, in contrast to the bulk that has an indirect band gap \cite{Bollinger:2001,Mak:2010,Splendiani:2010}. Access to the conduction band minimum (CBM) of the SL MoS$_2$ in ARPES is possible by doping with potassium. This is illustrated in Fig. \ref{fig:3} which shows a series of scans along the $\bar{M}-\bar{K}-\bar{\Gamma}$ and $\bar{A}-\bar{K}-\bar{A}'$ directions of the BZ for the clean surface and an increasing exposures to potassium. Overall, the expected strong electron doping is indeed observed: For a small potassium dose, all bands are shifted to higher binding energies and a weak photoemission intensity due to the conduction band minimum at $\bar{K}$ becomes observable (Fig. \ref{fig:3}(b) and (e)).  As the doping is increased, the CBM becomes populated, demonstrating the direct band gap of the material, as the VBM is also at $\bar{K}$ (Fig. \ref{fig:3}(c) and (f)). The CBM is found to be rather broad, in contrast to the VBM at $\bar{K}$, consistent with the out-of-plane character of these states \cite{Zhu:2011d,Cappelluti:2013}. We determine the gap energy to be  (1.39$\pm$0.05)~eV, substantially smaller than the value  of 1.88~eV determined by photoluminescence \cite{Mak:2010}.

Upon closer inspection, it becomes clear that potassium adsorption does not give rise to a simple rigid shift of the band structure to higher binding energies: While the VBM at $\bar{K}$ shifts by 1.39~eV - 1.56~eV =~-0.17~eV from the clean sample to the highly potassium-dosed situation, the maximum at $\bar{\Gamma}$ shifts by 1.70~eV - 2.03~eV =~-0.33~eV, such that the upper VB is severely distorted upon doping. This is again ascribed to the different orbital character near the high symmetry points, with the states at $\bar{\Gamma}$ being likely to show a stronger adsorbate-substrate interaction. Note that the same effect might also contribute to the small observed gap: The CBM states have a similar symmetry as the upper VB states at $\bar{\Gamma}$ and the observed gap for the potassium-dosed case is thus likely to be smaller than the gap of the pristine epitaxial SL MoS$_2$. 

In conclusion, we have studied the electronic structure of epitaxial single-layer MoS$_2$ on Au(111). We find this to give rise to sharp bands, in particular for the VBM near the $\bar{K}$ point and these bands are minimally affected by the presence of the substrate or moir\'e. This is very different from the $\pi$ band in epitaxial graphene and we ascribe this to the different orbital character of the bands (in-plane for MoS$_2$ versus out-of-plane for graphene). We directly observe the strong spin splitting of the upper VB and the size of this is in excellent agreement with theoretical predictions. Upon doping the layer with potassium, we are able to determine the band gap and find that the doping-induced shifts in energy are strongly dependent on the orbital character of the bands and hence $k$. This opens interesting possibilities to intentionally tune the band structure of SL MoS$_2$ and similar materials. The effect also needs to be taken into account when placing  SL MoS$_2$ between other materials or when doping it by an electric field.

We gratefully acknowledge financial support from the VILLUM foundation, The Danish Council for Independent Research, the Lundbeck Foundation,The Danish Strategic Research Council  (CAT-C) and Haldor Tops\o e A/S.


\end{document}